
\magnification=1200
\hoffset=.25truein
\hsize=6truein
\vsize=8.6truein
\nopagenumbers
\overfullrule=0pt
\pretolerance=10000
\tolerance=10000
\input epsf

\def\gtorder{\mathrel{\raise.3ex\hbox{$>$}\mkern-14mu
             \lower0.6ex\hbox{$\sim$}}}
\def\ltorder{\mathrel{\raise.3ex\hbox{$<$}\mkern-14mu
             \lower0.6ex\hbox{$\sim$}}}

\baselineskip=11pt
\centerline{\bf MONTE CARLO METHODS FOR NUCLEAR STRUCTURE}
\medskip
\centerline{STEVEN E. KOONIN}
\centerline{\it W. K. Kellogg Radiation Laboratory}
\centerline{\it California Institute of  Technology, Pasadena, CA 91125
USA}
\centerline{e-mail: sek@krl.caltech.edu}

\baselineskip=12pt plus1pt minus1pt
\noindent
{\bf 1. Introduction}
\vskip.14truein

This talk is about some new and powerful methods for dealing with the nuclear
shell model.  These Monte Carlo methods, which have been developed at
Caltech during the past five years, arise from a confluence of improvements
in both algorithms and computer power.  For selected observables, they allow
calculations much larger and more realistic than any possible by other
methods.  The applications to date demonstrate the power and potential of the
methods and the results in hand already offer a number of interesting
physical
insights, which I will describe to you.  In addition, the path is now clear
enough
to project with some certainty  further developments and
accomplishments in these matters during the next few years.

My presentation is organized as follows.  First, I will tell you what Shell
Model
Monte Carlo (SMMC) methods are capable of---what we can (and cannot)
calculate.  I would then like to give you some feeling for the general
strategy of
the calculations---how one deals with Hilbert spaces whose dimensions run to
billions or more.  My main focus will be on the application of these methods
to
problems of physical interest.  In particular, I will discuss complete {\it
pf}-shell
calculations of both the ground state and thermal properties of Fe-peak
nuclei.  I will then turn to two topics in collective motion:  the behavior
of a
hot, spinning rare-earth nucleus and of $\gamma$-soft nuclei near $A=124$.  I
will also show you how these methods can be used to calculate two-neutrino
double-beta decay rates.  Finally, I will offer some thoughts on how this
work will evolve over the next few years.  I will emphasize the basic ideas
and
physics results throughout; the development and details of the method can be
found in some of the early papers${}^1$, as well as in a forth-coming review
article${}^2$.

\vskip.24truein
\noindent
{\bf 2. What We Can Calculate}
\vskip.14truein

The SMMC methods I will describe are well suited to calculating thermal
averages of observables.  We cannot calculate the properties of any specific
state, except for the ground state (which is obtained by going to very low
temperature).  Although this precludes detailed spectroscopy, it is not as
limiting as it might seem.  Inclusive reactions and astrophysical
applications
require the properties of nuclei at finite temperatures, and the exact ground
state carries such interesting information as sum rules, pair correlations,
etc.

Within these thermal ensembles we can calculate averages of few-body
observables.  Most of the interesting physics can be had from the one- and
two-body density matrices, although the double beta-decay calculation
described in Section~8
requires a four-body observable.  The lack of explicit wavefunctions is not
too much of
a handicap, as the billions of amplitudes are of little interest in
themselves.
(After all, what experiment has ever measured a wavefunction!)

In addition to static observables, we can calculate information about the
strength functions for one-body operators:
$$
S_A(\omega)=Z^{-1} \sum_{if} e^{-\beta E_i}
\delta(\omega-E_f+E_i)
|\langle f|A|i\rangle|^2
\eqno(1)
$$
where $A$ is the operator of interest, $(i,f)$ are exact many-body
eigenstates
with energies $E_{(i,f)}$, $\beta$ is the inverse temperature, and $Z$ the
nuclear
partition function; spectral functions (where $A$ is an annihilation or
creation
operator) can also be calculated.  The strength functions are obtained from
the
Laplace transform of the corresponding imaginary-time response functions.
While the gross features of strength distributions are readily obtained, fine
detail is quite difficult, but this is often quite commensurate with the
experimental resolution.

SMMC results can be obtained with fully realistic shell model hamiltonians
(for
example, those derived from a $G$-matrix), although schematic interactions
have
been used as well.  Most of the results I will show you include all
configurations in one major shell, and multi-shell calculations are just
beginning.  Our confidence in the validity of the SMMC results (within the
quoted
uncertainties and the defined model) is based on careful comparisons with the
results
of more conventional methods, where the latter are feasible.

\vskip.24truein
\noindent
{\bf 3. General Strategy}
\vskip.14truein

Let me now give you some feel for the general strategy by which we
circumvent the combinatorial explosion of effort required in
conventional shell model methods.
As I mentioned, we consider thermal averages in the canonical (fixed-number)
ensemble at an inverse temperature $\beta$: $\langle A \rangle=
{\rm Tr}\,(e^{-\beta H} A)/{\rm Tr}\, e^{-\beta H}$.

The two-body interactions in the hamiltonian $H$ cause all of the trouble; if
$H$ were pure one-body with $N_s$ single-particle states, the trace
over all many-body states could be
readily evaluated by manipulations of $N_s \times N_s$ matrices.  The
``trick''
in SMMC is to transform the many-body problem into an infinite set of
one-body problems, each in a different external field.  The quantum mechanics
of
each of these is now quite tractable, but the price to be paid is the
necessity to
perform a weighted sum over all possible field configurations.

This latter task is handled by Monte Carlo methods, where only a statistical
sample of the most important field configurations is considered.  The
calculations are done on massively parallel computers, where each
computational node is tasked to produce and analyze a few field
configurations, and the final result is obtained by averaging over all nodes.
The
statistical uncertainty decreases as the square-root of the number of
samples;
typically several thousand yield the required precision.
In contrast to conventional diagonalization methods, the numerical effort in
SMMC is independent of the number of valence nucleons involved and scales
only as $N^3_s$.

\vskip.24truein
\noindent
{\bf 4. Ground States in the Fe Region}
\vskip.14truein

The first applications I will discuss are the properties of 28 nuclei
(even-even Ti,
Cr, Fe, Ni, and Zn isotopes, together with odd-odd $N=Z$ nuclides) in the Fe
region.${}^3$  The single-particle orbitals used were the $1p0f$-shell ($N_s
=20$)
and the hamiltonian was the Kuo-Brown interaction (KB3) with a tiny
monopole adjustment.  The calculations were done at $T=0.5~{\rm MeV}$,
which experience has shown to be sufficiently large so that observables
correspond essentially to ground-state properties.

\midinsert
\vbox{\baselineskip=10pt\parindent=0pt
\leftskip=.5truein\rightskip=.5truein
 Fig.~1: Upper panel: experimental and calculated mass defects.  Lower panel:
 Errors in the SMMC mass defects.  The horizontal lines show the average
(solid)
and rms (dashed) errors; a typical SMMC uncertainty is indicated.
}
\hskip.4truein
\vbox{\baselineskip=10pt\parindent=0pt
\leftskip=.5truein\rightskip=.5truein
Fig.~2: Calculated and experimental $E2$ strengths.  Open circles show the
experimental $0^+_1\rightarrow 2^+_1$ values,
while solid squares show the
experimental total values.
}
\endinsert

As shown in Fig.~1, the Coulomb-corrected mass defects are in excellent
agreement with experiment, the average error being $+0.45$~MeV (this is
consistent with the average internal excitation energy expected).  Figure~2
shows the calculated total $B(E2)$ strengths, using effective charges $(e_p,\
e_n)=(1.35,\ 0.35)$ that are consistent with conventional shell model
experience.  Comparisons with experimental values for only the lowest $2^+$
state are good, and the agreement with the total strengths, where available
from electron scattering, is excellent.

The individual $j$-shell occupations in the Fe isotopes are shown in Fig.~3.

The successive neutrons generally occupy the higher orbitals, although the
influence of the $f_{7/2}$ shell closure in ${}^{54} {\rm Fe}$ is evident, as
is the
effect of the neutron-proton interaction on the proton occupations.  To get
some
sense of the coherence in the ground states, one can consider the proton BCS
pairing gap, $\langle\Delta^\dagger_p\Delta_p\rangle$,  where
$\Delta^\dagger_p= \sum p^\dagger_{jm} p^\dagger_{j\bar m}$, the sum
being over all orbitals with $m>0$; the neutron gap is defined analogously.
Figure 4 shows the proton and neutron gaps for the Fe, Ni, and Zn isotopes
relative to those calculated for a Fermi gas with the same occupation
numbers.  The
proton pairing increases as neutrons are added, while the neutron pairing
clearly
shows the $f_{7/2}$ shell closure. A broader (and less model-dependent) view
of
pair correlations can be had by defining the operators $A^\dagger \equiv
[a^\dagger
\times a^\dagger ]_{JM}$ for each pair of orbitals, and then diagonalizing
the matrix
$\langle A^\dagger_{JM} A_{JM} \rangle$ for each $J$.  For $J=0$, the largest
eigenvalue
far exceeds the others, and defines the ``optimal'' pair content and
wavefunction.
As discussed in Section~7 below, channels with $J > 0$ can also show
meaningful coherence.

\midinsert
\hbox to \hsize{
\vtop{\baselineskip=10pt\parindent=0pt
\hsize=2.85truein
Fig.~3:  Calculated occupation numbers in the ground states of the Fe
isotopes.  Error bars are generally too small to be shown.
}
\hfill
\vtop{\baselineskip=10pt\parindent=0pt
\hsize=2.85truein
Fig.~4: Proton and neutron gaps (relative to the Fermi gas values) for the
Fe, Ni, and Zn isotopes.
}}
\endinsert

\midinsert
\vbox{\baselineskip=10pt\parindent=0pt
\leftskip=.5truein\rightskip=.5truein
Fig.~5: Experimental and calculated  ${\rm GT}_+$ strengths.  Discrepancies
for ${}^{48}$Ti and  ${}^{64}$Ni are likely due to deficiencies in the
model space used.
}
\endinsert

Particularly significant are the SMMC results for the total Gamow-Teller (GT)
strengths, $B({\rm GT}_\pm)=\langle G_\mp G_\pm\rangle$, with
$G_\pm=\sum\sigma t_\pm$
the usual
GT operator.  $B({\rm GT}_+)$ as measured in forward $(n,p)$ reactions is
typically only some 30\% of the independent particle estimate.  The Monte
Carlo results shown in Fig.~5 resolve this
discrepancy: the complete shell-model calculations systematically reproduce
the experimental values, provided the former are normalized by
$0.64=(1/1.25)^2$.  This is consistent with in-medium quenching of the axial
charge to $g_A=1$, as $\beta$-decay matrix elements are used to normalize the
experimental $(n,p)$ results.  A similar situation holds in the $sd$-shell.
These
data also show that  $B({\rm GT}_+)$ is proportional to the numbers of
valence
protons and valence neutron holes, so that the four $pf$-orbitals apparently
behave as one large entity.${}^4$

The extent to which the agreement between the data and the SMMC results
survives (or can be improved) with other hamiltonians or larger model spaces
will
be explored in the near future.

\vskip.24truein
\noindent
{\bf 5. Thermal Properties of Fe-Peak Nuclei}
\vskip.14truein

To investigate thermal properties, we have considered nuclei near Fe at
finite
temperature.   The calculations${}^5$ include the complete set of $1p0f$
states
interacting through the realistic Brown-Richter or Kuo-Brown Hamiltonians.

\midinsert
\hbox to \hsize{
\vtop{\baselineskip=10pt\parindent=0pt
\hsize=2.85truein
Fig.~6: Temperature dependence of various observables in  ${}^{54}$Fe.
Shown in the left-hand column are the internal energy, $U$, the heat capacity
$C$, and the level density parameter $a$.  In the right-hand column
are the neutron and proton pairing fields (values calculated in an
uncorrelated
Fermi gas are shown by the solid curves) and the moment of inertia,
 $I=\beta \langle J^2 \rangle/3$.
}
\hfill
\vtop{\baselineskip=10pt\parindent=0pt
\hsize=2.85truein
Fig.~7: Temperature dependence in  ${}^{54}$Fe of the total magnetic dipole
strength  $B(M1)$ (calculated using free-nucleon $g$-factors),
Gamow-Teller strength,
and isoscalar and isovector quadrupole
strengths.
}}
\endinsert

The calculated temperature dependence of various observables in ${}^{54}$Fe
is
shown in Fig.~6.  The internal energy $U$ increases steadily with increasing
temperature.  It shows an inflection point  around $T \approx 1.1$ MeV,
leading to a peak in the heat capacity, $C\equiv  dU/dT$.  The decrease in
$C$
for $T \gtorder 1.4$~MeV is due to the finite model space (Schottky effect);
the
limitation to only the $pf$-shell renders the calculations quantitatively
unreliable for temperatures above this value (internal energies $U\gtorder
15$ MeV). The same behavior is apparent in the level density parameter,
$a\equiv  C/2T$. The empirical value for $a$ is $A/(8~{\rm MeV}) =6.8~{\rm
MeV}^{-1}$, which is in good agreement  with the results for $T \approx
1.1$--1.5~MeV.

Also shown in Fig.~6 are the proton-proton and neutron-neutron BCS pairing
fields, $\langle\Delta^\dagger\Delta\rangle$. At low temperatures, the
pairing fields are significantly larger than those calculated for a
non-interacting
Fermi gas.  With increasing temperature, the pairing fields decrease,
approaching the Fermi gas values for $T\approx 1.5$~MeV and following  it
closely
for even higher temperatures. Associated with the breaking of  pairs is a
dramatic increase in the moment of inertia, $I$, for $T=1.0$--1.5~MeV. At
temperatures above 1.5~MeV, $I$ is in agreement with the rigid rotor value,
$10.7\hbar^2$/MeV; at even higher temperatures it  decreases linearly due to
the finite model space.

Some other static observables are shown in Fig.~7. The magnetic dipole
strength, $B(M1)$,
unquenches rapidly with heating near the transition temperature, but remains
significantly lower than the single-particle estimate ($41~\mu_N^2$) for
$T=1.3$--2~MeV, suggesting a persistent quenching at temperatures
above the like-nucleon depairing. This finding is supported by the
near-constancy of $B({\rm  GT}_+)$ for temperatures up to 2~MeV, as is often
assumed in astrophysical calculations.  Detailed study of pairing observables
shows that these behaviors are driven largely by the rapid vanishing of the
like-nucleon $J=0$ pairing near $T=1.1$~MeV and a peaking of the unlike
$J=1$ pairing near $T=2.25$~MeV.

\midinsert
\vbox{\baselineskip=10pt\parindent=0pt
\leftskip=.5truein\rightskip=.5truein
Fig.~8:  GT${}_+$ response functions (left) and corresponding strength
distributions
(right) at temperatures of roughly 1~MeV.  Data from  $(n, p)$ reactions
are shown as
histograms and the vertical lines indicate the centroids used in current
astrophysical
calculations.
}
\endinsert

The GT${}_+$ strength distributions $S(E)$ can be obtained as the
inverse Laplace transform of the
response function $R(\tau)=\langle G_-(\tau)G_+(0)\rangle$, where the GT
operators are
in the imaginary-time Heisenberg picture.  Figure 8 shows these quantities
for  ${}^{51}$V,
${}^{59}$Co, and ${}^{55}$Co.  Reasonable agreement is found with the $(n,p)$
data for the
first two nuclei (the calculations have not yet been folded with the
experimental
energy resolution).  That the
centroid in the third case is significantly higher than that predicted by
Fuller {\it et al.}
indicates that an important presupernova electron capture rate is likely
significantly
slower than the currently accepted value.

In a related study, SMMC calculations have been applied${}^6$ to test a
suggestion${}^7$ that the nuclear symmetry energy increases with temperature
due to a decrease in the nucleon effective mass.  If true, this would
decrease
supernova electron-capture rates.  To test this hypothesis, we performed SMMC
calculations of the differences in internal energies for several pairs of
even-even isobars at finite temperatures.  The calculations generally do not
support a
temperature-dependent increase in the symmetry energy.

Finite temperature SMMC calculations for other nuclei in this mass range show
behavior
similar to that of ${}^{54}$Fe; the qualitative features are also similar
when other
realistic interactions are used.  Results at temperatures above 1.5~MeV will
become
reliable only when two or more major shells are included in the calculations.
 The
extension of these studies to other interactions, heavier nuclei, and other
observables will allow a more thorough understanding of nuclear properties at
high excitation energies.

\vskip.24truein
\noindent
{\bf 6. ${}^{\bf 170}$Dy at Finite Temperature and Spin}
\vskip.14truein

The results of a calculation for ${}^{170}$Dy~${}^8$ demonstrate what SMMC
methods can bring to the description of heavier nuclei.  The protons occupied
the $Z=50$--82 shell while the neutrons were in the $N=82$--126 shell ($N_s$
was thus 32 and 44, respectively).  The nucleus ${}^{170}$Dy is of no
special interest physically, but as it is mid-shell in this model space (16
valence
protons and 22 valence neutrons), it is the most challenging (there are
some $10^{21}$ $m$-scheme determinants).  The hamiltonian was of the
conventional pairing plus quadrupole form.  Both grand-canonical (fixed
chemical potential) and canonical ensembles were used (and found to be quite
similar) and finite rotations were investigated by adding a cranking term
$-\omega J_z$ to the hamiltonian.

\midinsert
\hbox to \hsize{
\vtop{\baselineskip=10pt\parindent=0pt
\hsize=2.85truein
Fig.~9:  Grand-canonical observables for ${}^{170}$Dy at various cranking
frequencies and temperatures. Shown are the average sign
 $\langle\Phi\rangle$ (a smaller value indicates a numerically more difficult
calculation), the square of the isoscalar quadrupole moment  $\langle
Q^2\rangle$, the
energy  $\langle H\rangle$, the square of the angular momentum
 $\langle
J^2\rangle$, the dynamical moment of inertia
 $I_2=d\langle J_z\rangle/d\omega$, and the expectation value of the
pairing terms in the hamiltonian,  $-g\langle P^\dagger P\rangle$. Error bars
not shown are approximately the size of the symbols, and lines are
drawn to guide the eye.
}
\hfill
\vtop{\baselineskip=10pt\parindent=0pt
\hsize=2.85truein
\vskip 2.85truein
Fig.~10: Contours of the free energy in the polar-coordinate
 $\beta-\gamma$
plane for  ${}^{170}$Dy. Contours are shown at 0.3~MeV
intervals, with lighter shades indicating the more probable nuclear shapes.
}}
\endinsert

The systematics of the cranked system are shown in Fig.~9. At high
temperatures, the nucleus is unpaired and the moment of inertia decreases as
the system is cranked. However, for lower temperatures when the nucleons are
paired, the moment of inertia initially increases with $\omega$, but then
decreases at larger cranking frequencies as pairs break; the pairing gap
also decreases as a function of $\omega$. It is well known that
the moment of inertia is supressed by pairing and that initially ${\cal I}_2$
should increase with increasing $\omega$. Once the pairs have been broken,
the moment of inertia decreases. These features are evident in the figure.

In addition to simple observables, the nuclear shapes were calculated from
the
mass quadupole tensor of each Monte Carlo sample; the free energy can be
extracted from the distribution of these shapes.  Figure~10 shows the
temperature evolution of the free-energy surface.  At high temperatures, the
system is nearly spherical, whereas at lower temperatures, especially at
$T=0.3$~MeV,
there is a prolate minimum on the $\gamma=0$ axis.

Further systematic investigations of this sort are underway and expansion of
the single-particle basis to several major shells does not seem impossible.

\eject
\noindent
{\bf 7. Gamma-Soft Nuclei}
\vskip.14truein
The ground states of nuclei with $A\sim 124$ are expected to have large shape
fluctuations.  In geometrical terms, the potential energy surface has a
minimum
at finite $\beta$, independent of $\gamma$; in the Interacting Boson Model,
there is at $O(6)$ dynamical symmetry.  Recent SMMC work${}^9$ has provided
the first microscopic many-body description of this phenomenon.

Valence protons and neutrons were assumed to occupy the 50-82 shell (i.e.,
$N_s=32$).  Single particle energies were taken from a spherical Woods-Saxon
potential and the two-body interaction involved both monopole and
quadrupole pairing, as well as the usual $QQ$ term.  The parameters of the
Hamiltonian were adjusted to reproduce the experimental pairing energies and
excitation energies and $B(E2)$ values of the $2^+_1$ states.  Rotations were
investigated by cranking with  $-\omega J_z$.

\midinsert
\hbox to \hsize{
\vtop{\baselineskip=10pt\parindent=0pt
\hsize=2.85truein
Fig.~11: Cranking response of ${}^{124}$Xe and ${}^{128}$Te at temperatures
 $T = 0.5$ and 0.33~MeV.  Shown are the dynamical moment of inertia, $I$,
the mass quadrupole strength, the
proton pairing strength, and  $\langle J_z \rangle$.
}
\hfill
\vtop{\baselineskip=10pt\parindent=0pt
\hsize=2.85truein
Fig.~12: Similar to Fig.~10 for ${}^{124}$Xe and ${}^{128}$Te at temperatures
 $T = 1.0$, 0.5, and 0.33~MeV.
}}
\endinsert

The rotational responses of ${}^{124}$Xe  (4 valence protons and 20 valence
neutrons) and ${}^{128}$Te (2 protons and 26 neutrons) at two different
temperatures are shown in Fig.~11.  The $\omega=0$ inertia for both nuclei is
significantly lower than the rigid body value ($\sim 43 \hbar^2/{\rm MeV}$)
and increases with increasing rotation as the pairing decreases.  Peaks in
$I_2$
are associated with the onset of deformation as measured by $\langle
Q^2 \rangle$.  This suggests a band crossing associated with pair breaking
and
alignment, as is known to occur in ${}^{124}$Xe near spin $10\hbar$.  The
alignment is clearly seen in the behavior of $\langle J_z \rangle$ at the
lower
temperature, which shows a rapid increase after an initial moderate growth.
Both deformation and pairing decrease with increasing temperature.

Calculated free energies for ${}^{124}$Xe and ${}^{128}$Te are shown in
Fig.~12.
Both nuclei are essentially spherical at high temperature, but become
$\gamma$-soft at low temperature, with minima at $\beta \sim 0.15$ and 0.06,
respectively.  ${}^{128}$Te appears to be prolate, while ${}^{124}$Xe seems
to be
nearly $\gamma$-unstable.

A crude point of contact with the IBM can be had by calculating the numbers
of
correlated $J=0$ and $J=2$ pairs (i.e., excesses beyond the mean-field
values) at low temperature and
comparing them with the expected numbers of $S$ and $D$ bosons.  For
${}^{124}$Xe, the SMMC (IBM) results for protons are 0.85 (1.22) $S$-pairs
and 0.76
(0.78) $D$-pairs, where the IBM values correspond to the exact $O(6)$ limit.
For
neutron holes in the same nucleus, the corresponding values are 1.76 (3.67)
$S$-pairs and 2.14 (2.33) $D$-pairs.  For protons, the $D/S$ ratio of 0.89 is
not far from
the $O(6)$ value of 0.64, while for neutron holes, the $D/S$ of 1.21 is
intermediate between $O(6)$ and $SU(3)$ (where it is 1.64), as is consistent
with neutrons filling the middle of the shell.  Although the total numbers of
$S$ and $D$ pairs in the SMMC calculations (1.61 for protons and 3.8 for
neutron holes) are somewhat less
than the IBM values of 2 and 6, respectively, there is clear indication of
pair correlations at non-zero $J$.

\vskip.24truein
\noindent
{\bf 8. Double Beta Decay}
\vskip.14truein
The second-order weak process $(Z,A)\rightarrow(Z+2,A)+2e^-+2\bar\nu_e$
is an important ``background'' to searches for the lepton-number violating
neutrinoless mode, $(Z,A)\rightarrow(Z+2,A)+2e^-$. The calculation of the
nuclear
matrix elements for these two processes is a challenging problem in nuclear
structure, and has been done in a full $pf$ model space with conventional
methods only for  the lightest of
several candidates, ${}^{48}$Ca.

The required matrix element for $2\nu$ decay is
$$
M_{2\nu}=\sum_n
{\langle f|G_-|n\rangle\langle n|G_-|i\rangle\over
\Omega-E_n}
\eqno(2)
$$
Here, $(i,f)$ are the $0^+$ ground states of the parent and daughter nuclei,
$\Omega$ is the average of their energies, and the sum is over all $1^+$
states,
$n$, of the intermediate nucleus with energies $E_n$.  The difficulties in a
direct diagonalization approach involve
knowing the exact wavefunctions of the states involved, performing the sum
over all intermediate states, and the large cancellations that occur among
the
various terms in the sum.

To calculate ${M_{2\nu}}$ in very large model spaces with SMMC methods, one
considers the observable${}^{10}$
$$
F(\tau_1)={{\rm Tr}\,[e^{-(\beta-\tau-\tau_1)H}
G_+G_+ e^{-\tau H}G_-e^{-H\tau_1}G_-]\over
{\rm Tr}\,e^{-\beta H}}\;.
\eqno(3)
$$
If both $\beta$ and $\tau$ are
sufficiently large to filter out the parent and daughter ground states,
setting
$\tau_1=0$ leads to the usual closure matrix element, $M_c=\langle f|G_-G_-
|i\rangle$, while integrating over $\tau_1$ generates the required
intermediate-state energy denominator and hence leads to the exact
$M_{2\nu}$.

A first calculation to validate the SMMC method against direct
diagonalization
results has been performed for ${}^{48}$Ca. There is good agreement
between the SMMC and conventional results for the closure and exact matrix
elements;
cancellations make the matrix
element for ${}^{48}$Ca anomalously small, and hence the calculation
particularly demanding.  A similar calculation for ${}^{76}$Ge in the
$1p0f_{5/2}0g_{9/2}$ model space is
in progress. Of particular interest will be the sensitivity to the
effective interaction, the overlap of ${\rm GT}_+$ and ${\rm GT}_-$ strengths
in the intermediate nucleus, and the validity of both the closure
approximation
and the more sophisticated quasi-particle RPA.  Candidates for follow-on
calculations include ${}^{82}$Se, ${}^{96}$Zr, ${}^{100}$Mo,
${}^{128,130}$Te, and
${}^{136}$Xe.

\vskip.24truein
\noindent
{\bf 9. Summary and Outlook}
\vskip.14truein
I have presented a sampling of results from Shell Model Monte Carlo
calculations.  These
demonstrate both the power and limitations of the methods and the physical
insights they offer.  SMMC calculations, while not a panacea, clearly have
certain advantages over conventional shell model approaches, particularly for
properties of ground states or thermal ensembles.  Of the results I have
discussed, the most significant bear on the quenching of GT strength, on the
pairing structure, and on nuclear shapes.

With respect to the technical aspects of these calculations I note the
following:
\medskip
\item{1)}SMMC methods are computationally intensive.  However, computing
power
is becoming cheaper and more widely available at an astonishing rate.  It is
a
great advantage that these calculations can efficiently exploit loosely
connected
``farms'' of work-station-class machines.

\item{2)}We already have strong circumstantial evidence that center-of-mass
(CM)
motion is not a significant concern for many of the operators of interest
(the $E1$
operator being an outstanding exception).  This is
not too surprising at finite temperature, since the CM is only three degrees
of
freedom (far fewer than the internal dynamics).  Indeed, multi-shell
calculations have been initiated.

\item{3)}We lack the ability to treat odd-$A$ or odd-odd $N \neq Z$ systems
at
low temperatures because of ``sign'' problems in the Monte Carlo sampling.
Similar problems prevent spin projection, which would enable yrast
spectroscopy.  Work to circumvent these problems is clearly needed.

\item{4)}Otsuka and collaborators${}^{11}$ have recently proposed a hybrid
scheme
whereby SMMC methods are used to select a many-body basis, which is then
employed in a conventional diagonalization.  The sign problems alluded to
above are absent, and detailed spectroscopy is possible.  Test applications
to
boson problems have shown some promise, although the utility for realistic
fermion systems remains to be demonstrated.
\medskip

Additional physics results that should emerge in the next year or two
include:
more realistic electron capture rates in pre-supernova conditions, the
two-neutrino
double-beta decay matrix elements for ${}^{76}$Ge, and ${}^{128,130}$Te,
systematic
studies of rare earth nuclei at finite temperature and spin, studies to
improve
the effective interactions used, tests of such models as the IBM and RPA, and
predictions of nuclear properties far from $\beta$-stability. There are
undoubtedly
interesting applications beyond these, and suggestions are welcome.

\vskip.24truein
\noindent
{\bf 10. Acknowledgments}
\vskip.14truein
The results presented here are the collective accomplishments of D.~J.~Dean,
C.~W.~Johnson, G.~Lang, K.~Langanke, W.~E.~Ormand, P.~B.~Radha, and the
author at
Caltech and Y.~Alhassid at Yale.  We have benefited from interactions with
P.~Vogel, G.~Bertsch, T.~Kuo, A.~Poves, T.~Ressell, and J.~White.  The work
was supported in
part by the National Science Foundation (Grants PHY91-15574, PHY94-12818,
and PHY94-20470).  Computational cycles were provided by
the Caltech-based Concurrent Supercomputing Consortium, RIKEN, and the
Maui HPCC.

\vskip.24truein
\noindent
{\bf 11. References}
\vskip.14truein
\item{1.} C. W. Johnson {\it et al.} Phys. Rev. Lett. {\bf 69} (1992) 3157;
G. H. Lang {\it
et al.}, Phys. Rev. {\bf C48} (1993) 1518; Y. Alhassid {\it et al.}, Phys.
Rev. Lett.
{\bf 72} (1994) 613; W. E. Ormand {\it et al.} Phys. Rev. {\bf C49} (1994)
1422.

\item{2.} S. E. Koonin, D. J. Dean, and K. Langanke, to be published in
Physics
Reports.

\item{3.} D. J. Dean {\it et al.}, Phys. Rev. Lett. {\bf 72} (1994) 4066; K.
Langanke {\it et
al.}, Phys. Rev. {\bf C} (August, 1995).

\item{4.} S. E. Koonin and K. Langanke, Phys. Lett. {\bf B326} (1994) 5.

\item{5.} D. J. Dean {\it et al.}, Phys. Rev. Lett. {\bf 74} (1995) 2909; K.
Langanke {\it et al.},
to be submitted to Phys. Rev. C.

\item{6.} D. J. Dean {\it et al.}, Phys. Lett. B, in press.

\item{7.} P. Donati {\it et al.}, Phys. Rev. Lett. {\bf 72} (1994) 2835.

\item{8.} D. J. Dean {\it et al.}, Phys. Lett. B317 (1993) 275.

\item{9.} Y. Alhassid {\it et al.}, (July 1995), LANL preprint archive
nucl-th/9507029, submitted to Phys.
Rev. Lett.

\item{10.} P. B. Radha {\it et al.}, unpublished.

\item{11.} M. Honma, T. Mizusaki, and T. Otsuka, Phys. Rev. Lett. {\bf 75}
(1995) 1284.

\end